\documentclass[lettersize,journal]{IEEEtran}
\usepackage{amsmath,amsfonts}
\usepackage{algorithmic}
\usepackage{algorithm}
\usepackage{array}
\usepackage[caption=false,font=normalsize,labelfont=sf,textfont=sf]{subfig}
\usepackage{textcomp}
\usepackage{stfloats}
\usepackage{url}
\usepackage{verbatim}
\usepackage{graphicx}
\usepackage{cite}
\usepackage{tabularx,tabulary}
\usepackage{threeparttable}
\usepackage{booktabs}

\usepackage{cleveref}
\usepackage{graphicx}
\usepackage{caption}
\usepackage{tabularray}
\usepackage{threeparttable}
\begin{document}

\title{Reconfigurable Intelligent Surface-Enabled Gridless DoA Estimation System for NLoS Scenarios}

\author{Jiawen Yuan,~\IEEEmembership{Graduate Student Member,~IEEE,} Shaodan Ma, ~\IEEEmembership{Senior Member,~IEEE,} Gong Zhang,~\IEEEmembership{Member,~IEEE,}
and Henry Leung, ~\IEEEmembership{Fellow,~IEEE}
\thanks{Jiawen Yuan and Gong Zhang are with the Key Lab of Radar Imaging and Microwave Photonics, Nanjing University of Aeronautics and Astronautics, Nanjing 211100, China (e-mail: yuanjiawen@nuaa.edu.cn; gzhang@nuaa.edu.cn). Shaodan Ma is with the State Key Laboratory of Internet of Things for Smart City, University of Macau, Macao 999078, China (e-mail: shaodanma@um.edu.mo). Henry Leung is with the Department of Electrical and Software Engineering, University of Calgary, Calgary T2N 1N4, Canada (e-mail: leungh@ucalgary.ca).}}

\markboth{Journal of \LaTeX\ Class Files,~Vol.~14, No.~8, August~2021}%
{Shell \MakeLowercase{\textit{et al.}}: A Sample Article Using IEEEtran.cls for IEEE Journals}


\maketitle

\begin{abstract}
The conventional direction-of-arrival (DoA) estimation approaches are effective only when the line-of-sight (LoS) link is available. In non-line-of-sight (NLoS) scenarios, the spatial angle can not be captured and thus the DoA estimation performance would be significantly degraded. A novel reconfigurable intelligent surface (RIS)-enabled gridless DoA estimation system for NLoS scenarios is proposed to address this challenge, where the RIS establishes a virtual LoS link between the base station and the targets. To extract the statistics of the signal, the RIS-enabled signal model in the covariance domain is proposed. We estimate the noise variance by constraining the Frobenius norm of the measurement error matrix to enhance the robustness to noise. Additionally, we reconstruct the Hermitian Toeplitz matrix by addressing the atom norm minimization (ANM) problem on the covariance-based matrix. To reduce the computation, an efficient iterative approach is designed to solve the ANM problem via the alternating direction method of multipliers. Numerical experiments validate the superiority of the proposed system over the benchmark in terms of computational efficiency and multi-source DoA estimation precision.
\end{abstract}


\section{Introduction}
\IEEEPARstart{D}{irection-of-Arrival} (DoA) estimation, which aims to mine angle information from the active or passive target, has drawn tremendous interest in the past years, especially in the realms of communication, radar, and Internet-of-Things \cite{ref1}-\cite{ref4}. Extensive literature based on DoA estimation has been proposed
, including a series of celebrated DoA algorithms based on the compressed sensing (CS) framework \cite{ref5}, \cite{ref6}. However, the aforementioned approaches are only effectively applied in the line-of-sight (LoS) scenario and cannot be easily extended to non-line-of-sight (NLoS) environments, such as urban settings with numerous potential obstacles.

Considering the complication of electromagnetic wave propagation in NLoS scenarios, the DOA estimation problem becomes more challenging. Specifically, since the propagation path in NLoS is longer than that in LoS, the intensity of the electromagnetic wave changes, which in turn imposes a requirement on the robustness of the DOA estimation algorithm. Thus, reconfigurable intelligent surface (RIS) \cite{ref7}, \cite{ref8} has recently been proposed as a ground-breaking candidate technology because of its capability to reconfigure wireless channels through intelligent signal reflections. This provides a fresh perspective to produce a virtual LoS link from the base station (BS), potentially enabling target localization in NLoS scenarios. A novel IRS-self-sensing architecture combining the IRS controller and dedicated sensors is proposed in \cite{ref9}. However, it can only be applied as a single target. In [10], the RIS-assisted system for the gridless DOA estimation is proposed for time domain. This leads to the following proposal of an atomic norm-based estimate approach in [11] for the scenario of unmanned aerial vehicle swarms employing RIS. Both algorithms exploit the target sparsity and apply a semidefinite programming (SDP) method to obtain DoAs, which achieve satisfactory estimation accuracy compared with traditional CS methods.
However, these methods exploit information from the time domain, which results in sensitivity to noise and establishes unreliable estimates in non-ideal conditions.

To enhance robustness to noise and obtain high-precision estimation, we concentrate on the multi-target active detection and propose a RIS-enabled gridless DoA estimation system for NLoS scenarios. It is divided into two parts, the former is denoising based on the Frobenius norm minimization criterion of the measurement error matrix, and the last is the fast covariance-based DoA estimation via the alternating direction method of multipliers (ADMM). 
Our main contributions are summarized as follows: 1) We propose a RIS-enabled signal model in the covariance domain to fully excavate the statistics information. 2) Based on the Frobenius norm minimization criterion of the measurement error matrix, we propose an estimation method for the noise variance. This provides a prerequisite for the subsequent improvement of the noise suppression performance. 3) We obtain the Hermitian Toeplitz matrix to estimate the multi-source DoAs via atomic norm minimization (ANM), which can overcome the basis mismatch and enhance the estimation accuracy under the gridless CS framework. 4) By applying the ADMM technique, we provide a fast approach to reduce the computational cost of the matrix reconstruction based on the ANM problem.
\section{Signal Model in the Time Domain}
\begin{figure}[!htb]
\centering
\includegraphics[width=2.527in]{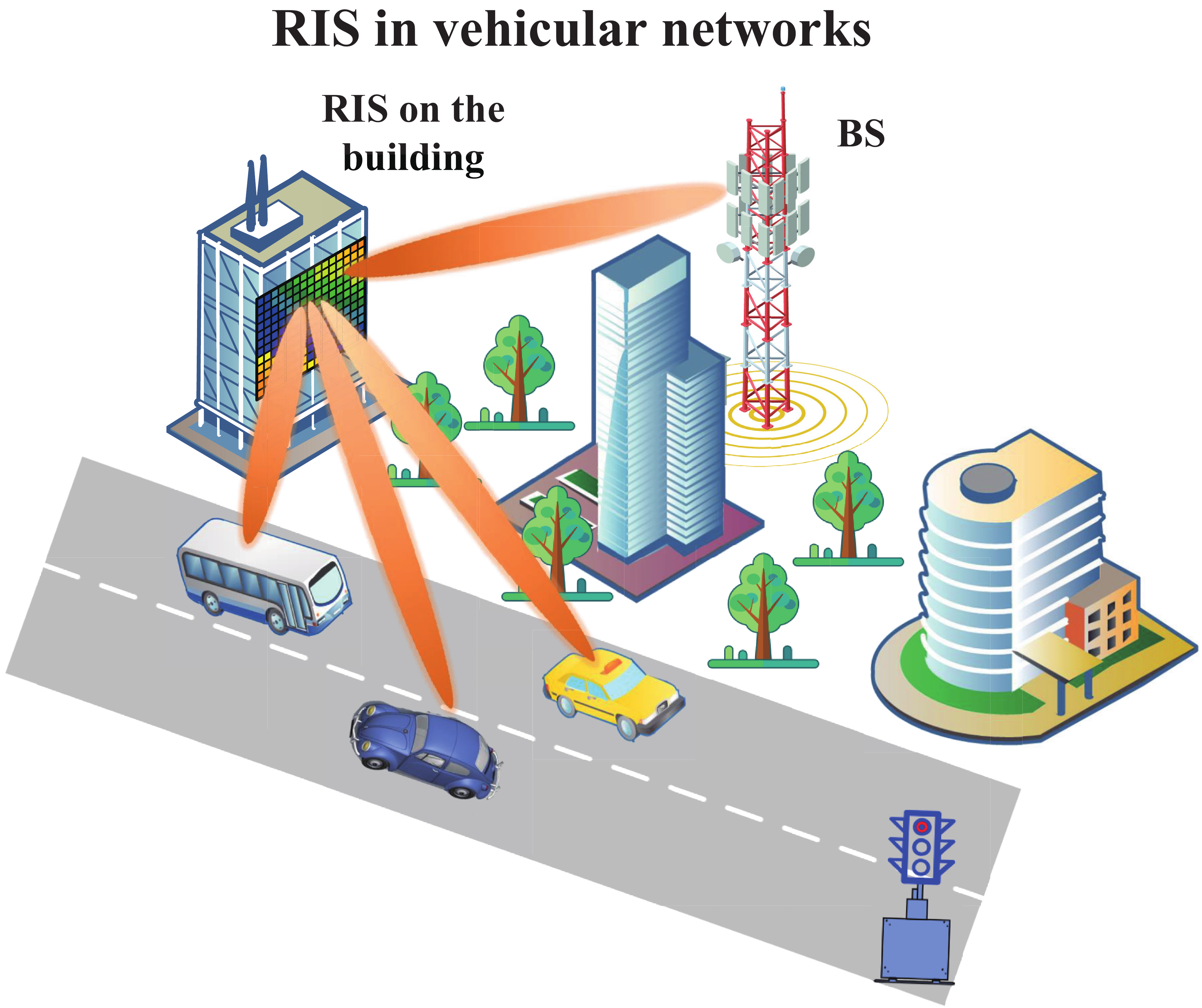}
\caption{RIS-enabled sensing scenario in vehicular networks.}
\label{fig:RIS1}
\end{figure}

This paper proposes a RIS-enabled active detection system to estimate the DoAs of the signal sources through NLoS propagation, with multiple omnidirectional antennas as the receiver. To visualize the potential applications of the system, we describe the RIS-enabled sensing scenario in vehicular networks in Fig.\ref{fig:RIS1}, including one BS, one RIS, and multiple vehicles. Note that the system is predicated on a 1D DoA estimation assumption, which is readily expanded to 2D DoA estimation problems. Consider that the BS and RIS are arranged linearly, where the number of elements is $M$ and $N$ respectively. Also, the position separation of RIS from $K$ sources and BS is assumed to be situated in the far field.

The incident signal for $n$-element RIS $x_{\mathrm{in},n}(t)$ impinged by $K$ narrowband far-field sources can be expressed as
\begin{equation} \label{RIS:eq1}
x_{\mathrm{in}, n}(t)=\sum_{k=1}^{K} s_k(t) e^{j \frac{2 \pi}{\lambda} p_n \sin \theta_k},
\end{equation}
where $s_k(t)$ and $\theta_{k}$ denote the signal and DoA actively transmitted by $k$-th target, respectively. $p_{n}$ represents the position of $n$-th element RIS with an interspacing half-wave length $\lambda$/2. Assume that there is no hardware deficiency such as circuit nonlinearity and phase noise, the reflected signal by $n$-th element RIS can be expressed by
\begin{equation} \label{RIS:eq2}
x_{\mathrm{out}, n}(t)=G_n(t) e^{j \phi_n(t)} \sum_{k=1}^K s_k(t) e^{j \frac{2 \pi}{\lambda} p_n \sin \theta_k},
\end{equation}
where $G_n(t)$ and $e^{j \phi_n(t)}$ denote the reflection amplitude and phase at the $n$-th RIS element, respectively. Then the observed data of $m$-th receiving antenna $y_m(t)$ is defined as
\begin{equation}\label{RIS:eq3}
\begin{aligned}
y_m(t)=& \sum_{n=1}^N x_{out, n}(t) e^{j \frac{2 \pi}{\lambda}\left(p_n \sin \alpha+q_m \sin \beta\right)} \\
=& \sum_{n=1}^N \sum_{k=1}^K\!\!\left[\!\!\begin{array}{l}
G_n(t) e^{j \phi_n(t)} e^{\frac{j \pi}{\lambda} p_n \sin \theta_k} \\
\times e^{j \frac{2 \pi}{\lambda}\left(p_n \sin \alpha+q_m \sin \beta\right)}
\end{array}\!\!\right]\!\! s_k+v_m(t),
\end{aligned}
\end{equation}
where $\alpha$ and $\beta$ are the direction-of-departure (DoD) and DoA of RIS and BS respectively.  $v_m(t)$ and $q_{m}$ denote the additive Gaussian noise and the localization of $m$-th antenna, respectively. We divide time domain into $L$ time plots with $T$ period and set $s_k(lT) \approx s_k((l+1)T) \approx s_k (l=1,\cdots,L-1)$ under the narrowband supposition. Hence the discretized data at the $l$-th time slot for $m$-th receiving antenna can be written as
\begin{equation} \label{RIS:eq4}
\begin{aligned}
y_m(l T)\!=\!\sum_{n=1}^N \sum_{k=1}^K\!\left[\begin{array}{l}
G_n(l T) e^{j \phi_n(l T)} e^{j \frac{2 \pi}{\lambda} p_n \sin \theta_k} \\
\times e^{j \frac{2 \pi}{\lambda}\left(p_n \sin \alpha+q_m \sin \beta\right)}
\end{array}\!\!\right] s_k+v_m(l T).
\end{aligned}
\end{equation}
Furthermore, we sort each value $y_m(lT)$ into a matrix-wise form, resulting in the observation matrix $\mathbf{Y} \in \mathbb{C}^{L \times M}$ of $M$ receiving antennas as follows
\begin{equation} \label{RIS:eq5}
\begin{split}
\mathbf{Y}=&\!\left[\mathbf{y}_1, \ldots, \mathbf{y}_M\right]=\!\left[\!\!\!\begin{array}{cccc}
y_1(T) & \!y_2(T) & \!\!\cdots & \!\!y_M(T) \\
y_1(2T) & \!y_2(2T) & \!\!\cdots & \!\!y_M(2T) \\
\vdots & \!\vdots & \!\!\ddots & \!\!\vdots \\
y_1(LT) & \!y_2(LT) & \!\!\cdots & \!\!y_M(LT)
\end{array}\!\!\!\right] \\
=&\mathbf{B}^{\rm{T}} \mathbf{A}(\theta) \mathbf{s} e^{j \frac{2 \pi \sin \beta}{\lambda} \mathbf{q}^{\rm{T}}}+\mathbf{V},
\end{split}
\end{equation}
where $\mathbf{s}=\left[s_1, \ldots, s_K\right]^{\rm{T}} \in \mathbb{C}^{K}$ and $\mathbf{q}=\left[q_1,\ldots,q_M\right]^{\rm{T}}\in \mathbb{C}^{M}$. $\mathbf{B}=[\mathbf{b}(1), \ldots, \mathbf{b}(L)] \in \mathbb{C}^{N \times L}$, $\mathbf{A}(\theta)=\left[\mathbf{a}\left(\theta_1\right), \ldots, \mathbf{a}\left(\theta_K\right)\right] \in \mathbb{C}^{N \times K}$, and $\mathbf{V}=\left[\mathbf{v}_1, \ldots, \mathbf{v}_M\right] \in \mathbb{C}^{L \times M}$ denote the RIS-enabled measurement matrix, array manifold matrix, and noise matrix, respectively. The column vectors of the above matrices are defined as $\mathbf{b}(l) = e^{j \frac{2 \pi \sin \alpha}{\lambda} \mathbf{p}} \odot \mathbf{g_{ \boldsymbol{\phi}}}(l)$, $\mathbf{a}\left(\theta_k\right)=e^{j \frac{2 \pi \sin \theta_{k}}{\lambda} \mathbf{p}}$, and $\mathbf{v}_m=\left[v_m(T),\ldots,v_m(LT)\right]^{\rm{T}}$ respectively, where $\mathbf{p}=\left[p_1,\ldots,p_N\right]^{\rm{T}}$ and $\mathbf{g_{ \boldsymbol{\phi}}}(l)=\left[G_1(lT)e^{j\phi_1(lT)},\ldots,G_N(lT)e^{j\phi_N(lT)}\right]^{\rm{T}}$. $\left(\cdot\right)^{\rm{T}}$ and $\odot$ are transpose operator and Kronecker product, respectively.

\section{RIS-Enabled Gridless DoA Estimation System \\ in the Covariance domain}
In this section, we demonstrate the RIS-enabled gridless DoA estimation framework from the perspective of the covariance domain.

To begin with, we define the covariance matrix of $\mathbf{Y}$ as follows
\begin{equation}  \label{RIS:eq6}
\begin{split}
\mathbf{R}_{\mathbf{\!Y}}\!\!=& \mathrm{E}\!\left\{\mathbf{Y} \!\mathbf{Y}^{\rm{H}}\!\right\}\!\!=\!\!M \mathbf{B}^{\rm{T}} \!\mathbf{A}(\theta) \mathrm{E}\!\left\{\mathbf{s s}^{\rm{H}}\!\right\}\! \mathbf{A\!}^{\!\rm{H}}(\theta) \mathbf{B}^{*}\!\!+\!\!\mathrm{E}\!\left\{\mathbf{V} \mathbf{V}^{\rm{H}}\!\right\} \\
=& M \mathbf{B}^{\rm{T}} \mathbf{A}(\theta) \mathbf{R}_\mathbf{s} \mathbf{A}^{\rm{H}}(\theta) \mathbf{B}^*+\sigma_0 \mathbf{I}_L \\
=& M \mathbf{B}^{\rm{T}} \mathbf{R} \mathbf{B}^*+\sigma_0 \mathbf{I}_L = \mathrm{E}\left\{\mathbf{c c}^{\rm{H}}\right\}+\sigma_0 \mathbf{I}_L,
\end{split}
\end{equation}
where $\mathbf{R}_{\mathbf{s}}=\mathrm{E}\!\left\{\mathbf{s s}^{\rm{H}}\right\}$ and $\mathbf{R}=\mathbf{A}(\theta) \mathbf{R}_\mathbf{s} \mathbf{A}^{\rm{H}}(\theta)$ denote the signal covariance matrix and the RIS-enabled covariance matrix without noise interference, respectively. $\mathbf{c}\!=\!\sqrt{M} \mathbf{B}^{\rm{T}} \mathbf{A}(\theta) \mathbf{s}$ is the vector based on matrix decomposition. $\sigma_0$ is the variance of Gaussian white noise and $\mathbf{I}_L$ is the  $L$-dimensional unitary matrix. $\mathrm{E}\left(\cdot\right)$, $\left(\cdot\right)^{\rm{*}}$, and $\left(\cdot\right)^{\rm{H}}$ stand for the operator of mathematical expectation, conjugate, and conjugate transpose, respectively. Due to the interference of the noise, it is worth denoising to have a better estimation performance. Thus, the novel RIS-enabled gridless DoA estimation method is mainly divided into two stages, the first stage is to estimate the power of noise, and the second stage concentrates on the angle of sources.


\subsection{Noise Variance Estimation in the Covariance Domain}
Since the accuracy of the noise power estimate has a significant impact on the subsequent the covariance-based matrix reconstruction, we will minimize the Frobenius norm of the measurement error matrix to estimate the noise power. The detailed optimization problem is given as follows:
\begin{equation}\label{RIS:eq7}
\begin{aligned}
\left\{\hat{\mathbf{c}}, \hat{\sigma}_0\right\} = & \underset{\mathbf{c}, \sigma_0}{\arg \min } \left\|\Delta \mathbf{R}_{\mathbf{Y}}\right\|_F^2\\
&\text { s.t. } \Delta \mathbf{R}_{\mathbf{Y}}=\widehat{\mathbf{R}}_{\mathbf{Y}}-\mathrm{E}\left\{\mathbf{c c}^{\rm{H}}\right\}-\sigma_0 \mathbf{I}_L,
\end{aligned}
\end{equation}
where $\left\|\cdot\right\|_F$ denotes the Frobenius norm. $\Delta \mathbf{R}_{\mathbf{Y}} = \mathbf{R}_{\mathbf{Y}} - \widehat{\mathbf{R}}_{\mathbf{Y}}$ represents the measurement error matrix, where $\widehat{\mathbf{R}}_{\mathbf{Y}}$ is the a discretized sample of $\mathbf{R}_{\mathbf{Y}}$. Firstly, we expand $\left\|\Delta \mathbf{R}_{\mathbf{Y}}\right\|_F^2$ as
\begin{equation}\label{RIS:eq8}
\begin{split}
\!\!\!\!\!\!\left\|\Delta \mathbf{R}_{\mathbf{Y}}\right\|_F^2\!=& \!\left\|\widehat{\mathbf{R}}_{\mathbf{Y}}-\mathrm{E}\left\{\mathbf{c c}^{\rm{H}}\right\}-\sigma_0 \mathbf{I}_L\right\|_F^2 \\
=& \operatorname{tr}\!\left\{\!\!\left(\!\widehat{\mathbf{R}}_{\mathbf{Y}}\!-\!\mathrm{E}\!\left\{\!\mathbf{c c}^{\rm{H}}\!\right\}\!-\!\sigma_0 \mathbf{I}_L\!\right)\!\!\!\left(\!\widehat{\mathbf{R}}_{\mathbf{Y}}\!-\!\mathrm{E}\!\left\{\!\mathbf{c c}^{\rm{H}}\!\right\}\!-\!\sigma_0 \mathbf{I}_L\!\right)^{\rm{\!H}}\!\right\} \\
=& \operatorname{tr}\left\{\widehat{\mathbf{R}}_{\mathbf{Y}}^2\!\right\}-2\mathrm{E}\left\{\mathbf{c}^{\rm{H}}\mathbf{c}\right\}\operatorname{tr}\left\{\widehat{\mathbf{R}}_{\mathbf{Y}}\right\}-2\sigma_0 \operatorname{tr}\left\{\widehat{\mathbf{R}}_{\mathbf{Y}}\right\}\\
&+ 2 \sigma_0 \mathrm{E}\left\{\mathbf{c}^{\rm{H}}\mathbf{c}\right\}+\mathrm{E}^2\left\{\mathbf{c}^{\rm{H}}\mathbf{c}\right\}+L \sigma_0^2.
\end{split}
\end{equation}
Then, we focus on the partial derivative of $\left\|\Delta \mathbf{R}_{\mathbf{Y}}\right\|_F^2$ as follows
\begin{subequations} \label{RIS:eq9}
\begin{alignat}{1}
\!\!\frac{\partial \left\|\Delta \mathbf{R}_{\mathbf{Y}}\right\|_F^2}{\partial \sigma_0} \!=& \frac{ 2 \partial \sigma_0 \mathrm{E}\left\{\mathbf{c}^{\rm{H}}\mathbf{c}\right\} - 2 \partial \operatorname{tr}\left\{\sigma_0 \widehat{\mathbf{R}}_{\mathbf{Y}}\right\}+\partial\left\{L \sigma_0^2\right\}}{\partial \sigma_0} \notag\\
\!=& -2 \operatorname{tr}\left\{\widehat{\mathbf{R}}_{\mathbf{Y}}\right\}+2 \mathrm{E}\left\{\mathbf{c}^{\rm{H}} \mathbf{c}\right\}+2 L \sigma_0 \label{RIS:eq9a},\\
\!\!\frac{\partial \left\|\Delta \widehat{\mathbf{R}}_{\mathbf{Y}}\right\|_F^2}{\partial \mathbf{c}^*} \!=& \frac{2 \partial \operatorname{tr}\!\left\{\!\left(\sigma_0 \mathbf{I}_L \!-\! \widehat{\mathbf{R}}_{\mathbf{Y}} \right)\!\mathrm{E}\left\{\mathbf{c} \mathbf{c}^{\rm{H}}\right\}\!\right\}\!+\!\partial \! \left\{\mathrm{E}^2\!\left\{\mathbf{c}^{\rm{H}}\mathbf{c}\right\} \right\}\!}{\partial \mathbf{c}^*} \notag\\
\!=& 2 \sigma_0 \mathbf{c}-2 \widehat{\mathbf{R}}_{\mathbf{Y}} \mathbf{c}+2 \mathrm{E}\left\{\mathbf{c}^{\rm{H}} \mathbf{c}\right\} \mathbf{c}. \label{RIS:eq9b}
\end{alignat}
\end{subequations}
Setting $(\ref{RIS:eq9a})$ and $(\ref{RIS:eq9b})$ to be zero, the coupled parameters can be obtained as
\begin{subequations} \label{RIS:eq10}
\begin{alignat}{1}
\sigma_0=\frac{1}{L} \operatorname{tr}\left\{\widehat{\mathbf{R}}_{\mathbf{Y}}-\mathrm{E}\left\{\mathbf{c c}^{\rm{H}}\right\}\right\}, \label{RIS:eq10a}\\
\left(\widehat{\mathbf{R}}_{\mathbf{Y}}-\sigma_0 \mathbf{I}_L\right) \mathbf{c}=\mathrm{E}\left\{\mathbf{c c}^{\rm{H}}\right\} \mathbf{c}. \label{RIS:eq10b}
\end{alignat}
\end{subequations}
Perform eigenvalue decomposition on $(\ref{RIS:eq10b})$ and let $\widehat{\mathbf{R}}_{\mathbf{Y}}-\sigma_0 \mathbf{I}_L=\mathbf{U} \mathbf{\Sigma} \mathbf{U}^{\rm{H}}$, the estimation of $\mathbf{c}$ is
\begin{equation}\label{RIS:eq11}
\mathbf{c} = \mathbf{U}(:,1) \mathbf{\Sigma}_{1,1}^{\frac{1}{2}},
\end{equation}
with the value $\mathbf{\Sigma}_{1,1}$ in the first row and first column of $\mathbf{\Sigma}$ and the vector $\mathbf{U}(:,1)$ in the first column of $\mathbf{U}$.

It is noticed that the optimal solutions corresponding to the $\mathbf{c}$ and $\mathbf{\sigma}_0$ are solved in such a mutually iterative manner. The details are shown in Algorithm 1.

\begin{table}[!h]
\normalsize
\centering
\begin{tabular}{p{0.05cm}p{8cm}}
\toprule
   \multicolumn{2}{l}{$\bf{Algorithm~1}$ Estimation Method of Noise Variance} \\
\midrule
1. & \emph{Input:} The observation matrix $\mathbf{Y}$, the RIS-enabled measurement matrix $\mathbf{B}$, the RIS element location $\mathbf{p}$, the receiving antenna location $\mathbf{q}$, DoD $\alpha$, DoA $\beta$, the maximum number of iterations $I_1$, and the stagnation threshold $\epsilon_1$;\\
2. & \emph{Initialization:} let \emph{i} = $\sigma_0^0$ = 0, $\mathbf{c}^0=\mathbf{0}$, and the sample covariance matrix $\widehat{\mathbf{R}}_{\mathbf{Y}}$ calculated via $(\ref{RIS:eq6})$;\\
3. & $\mathbf{Repeat}$\\
& $\qquad$ $i=i+1$; \\
& $\qquad$ Update the vector $\mathbf{c}^i$ in closed form $(\ref{RIS:eq11})$;\\
& $\qquad$ Update the parameter $\sigma_0^i$ in closed form $(\ref{RIS:eq10a})$;\\
& $\mathbf{Until} \quad i>I_1$ or $\left|\sigma_0^{i}-\sigma_0^{i-1}\right|/\left|\sigma_0^{i-1}\right|<\epsilon_1$;\\
4. & \emph{Output:} estimated noise variance $\widehat{\sigma}_0=\sigma_0^{i}$.\\
\bottomrule
\end{tabular}
\end{table}

\subsection{RIS-Enabled DoA Estimation Method Based on ANM}
With the obtained $\widehat{\sigma}$ by Algorithm 1, the sample RIS-enabled covariance matrix $\widehat{\mathbf{R}}$ without noise nuisance can be expressed as
\begin{equation}\label{RIS:eq12}
\widehat{\mathbf{R}}=\left(\mathbf{B}^{\rm{T}}\right)^{\dagger}\left(\widehat{\mathbf{R}}_{\mathbf{Y}}-\hat{\sigma}_0 \mathbf{I}_L\right)\left(\mathbf{B}^*\right)^{\dagger}=\mathbf{A}(\theta) \widehat{\mathbf{R}}_{\mathrm{s}} \mathbf{A}^{\rm{H}}(\theta),
\end{equation}
where $\widehat{\mathbf{R}}_{\mathbf{s}}$ and $(\cdot)^{\dagger}$ are the signal-based sample covariance matrix and Moore–Penrose pseudoinverse, respectively. Thus, we give a new linear combination of $\mathbf{R}$ as
\begin{equation}\label{RIS:eq13}
\begin{split}
\mathbf{R} =& \mathbf{A}(\theta) \mathbf{R}_{\mathrm{s}} \mathbf{A}^{\rm{H}}(\theta)=\mathbf{A}(\theta) \mathbf{D} \\
=& \sum_{k=1}^K \mathbf{a}\left(\theta_k\right) \mathbf{d}_k^r=\sum_{k=1}^K \gamma_k \mathbf{a}\left(\theta_k\right) \overline{\mathbf{d}}_k^r,
\end{split}
\end{equation}
where $\mathbf{d}_k^r$ represents the $k$-th row vector of $\mathbf{D}=\mathbf{R}_{\mathrm{s}} \mathbf{A}^{\rm{H}}(\theta)$. $\overline{\mathbf{d}}_k^r = \mathbf{d}_k^r/\left\|\mathbf{d}_k^r\right\|_2$ satisfies $\left\|\overline{\mathbf{d}}_k^r\right\|_2=1$ and $\left\|\cdot\right\|_2$ denotes the $l_2$ norm. Note that we define a new atom set:
\begin{equation}\label{RIS:eq14}
\mathcal{A}\!=\!\left\{\mathbf{a}(\theta) \overline{\mathbf{d}}^r \mid \theta \!\in\!\left[-\frac{\pi}{2}, \frac{\pi}{2}\right), \overline{\mathbf{d}}^r \!\in\! \mathbb{C}^{1 \times N},\left\|\overline{\mathbf{d}}^r\right\|_2^2\!=\!1\right\}.
\end{equation}
Inspired by the atomic norm, we define the atom norm of $\mathbf{R}$ as
\begin{equation}\label{RIS:eq15}
\left\|\mathbf{R}\right\|_{\mathcal{A}}=\inf \left\{\sum_k \gamma_k \bigg| \mathbf{R}=\sum_k \gamma_k \mathbf{a}\left(\theta_k\right) \overline{\mathbf{d}}_k^r, \mathbf{a}\left(\theta_k\right) \overline{\mathbf{d}}_k^r \in \mathcal{A}\right\}.
\end{equation}
Thus, the optimal covariance matrix $\mathbf{R}$ can be recovered by solving the following ANM problem:
\begin{equation}\label{RIS:eq16}
\min _{\mathbf{R}}\|\mathbf{R}\|_{\mathcal{A}} \quad \text { s.t. }\left\|\widehat{\mathbf{R}} - \mathbf{R}\right\|_F^2 \leq \varepsilon.
\end{equation}
The minimization problem ($\ref{RIS:eq16}$) can be converted into an SDP problem:
\begin{equation}\label{RIS:eq17}
\begin{gathered}
\underset{\boldsymbol{\mu},\mathbf{W},\mathbf{R}}{\min} \quad \!\!\!\!\! \operatorname{tr}(\mathbf{T}(\boldsymbol{\mu})) + \operatorname{tr}(\mathbf{W})\\
\text { s.t. }\left[\!\begin{array}{cc}
\mathbf{W} & \mathbf{R}^{\mathrm{H}} \\
\mathbf{R} & \mathbf{T}(\boldsymbol{\mu})
\end{array}\!\right] \succeq 0,\left\|\mathbf{R}- \widehat{\mathbf{R}} \right\|_F^2 \leq \varepsilon,
\end{gathered}
\end{equation}
where $\mathbf{T}\left(\boldsymbol{\mu}\right) \in \mathbb{C}^{N \times N}$ denotes the Hermitian Toeplitz matrix controlled by $\boldsymbol{\mu}=\left[\mu_1,\dots,\mu_N\right]$. $\mathbf{W}$ and $\epsilon$ are the Hermitian matrix and the measurement error bound, respectively. Due to the convex property, we can get the solution in Lasso form by CVX toolbox \cite{ref12}
\begin{equation}\label{RIS:eq18}
\begin{split}
\underset{\boldsymbol{\mu},\mathbf{W},\mathbf{R}}{\min} \quad \!\!\!\!\!\! &\operatorname{tr}(\mathbf{T}(\boldsymbol{\mu})) + \operatorname{tr}(\mathbf{W}) + \gamma  \left\|\mathbf{R}- \widehat{\mathbf{R}} \right\|_F^2\\
&\text { s.t. }\left[\begin{array}{cc}
\mathbf{W} & \mathbf{R}^{\mathrm{H}} \\
\mathbf{R} & \mathbf{T}(\boldsymbol{\mu})
\end{array}\!\right] \succeq 0.
\end{split}
\end{equation}
where $\gamma$ is the regularization parameter. However, the application of the CVX toolbox is time-consuming, and fast algorithms need to be considered.

Therefore, we express our problem in a proper manner for ADMM \cite{ref13}, ($\ref{RIS:eq18}$) is rewritten as
\begin{equation}\label{RIS:eq19}
\begin{split}
\underset{\boldsymbol{\mu},\mathbf{W},\mathbf{R}}{\min} \quad \!\!\!\!\!\! &\operatorname{tr}(\mathbf{T}(\boldsymbol{\mu})) + \operatorname{tr}(\mathbf{W}) + \gamma  \left\|\mathbf{R}- \widehat{\mathbf{R}} \right\|_F^2\\
&\text { s.t. } \mathbf{Z} = \left[\begin{array}{cc}
\mathbf{W} & \mathbf{R}^{\mathrm{H}} \\
\mathbf{R} & \mathbf{T}(\boldsymbol{\mu})
\end{array}\!\right], \mathbf{Z} \succeq 0.
\end{split}
\end{equation}
and dualize the equality constraint via an Augmented Lagrangian:
\begin{equation}\label{RIS:eq20}
\begin{aligned}
& \!\!\mathcal{L}(\boldsymbol{\mu}, \mathbf{W}, \mathbf{R}, \boldsymbol{\Pi}, \mathbf{Z})=\operatorname{tr}(\mathbf{T}(\boldsymbol{\mu}))+\operatorname{tr}(\mathbf{W})+\gamma\|\mathbf{R}-\widehat{\mathbf{R}}\|_F^2 \\
& \!\!+\!\left\langle\boldsymbol{\Pi}, \mathbf{Z}\!-\!\left[\begin{array}{cc}
\mathbf{W} & \mathbf{R}^{\mathrm{H}} \\
\mathbf{R} & \mathbf{T}(\boldsymbol{\mu})
\end{array}\right]\right\rangle \!+\!\frac{\tau}{2}\left\|\mathbf{Z}\!-\!\left[\begin{array}{cc}
\mathbf{W} & \mathbf{R}^{\mathrm{H}} \\
\mathbf{R} & \mathbf{T}(\boldsymbol{\mu})
\end{array}\right]\right\|_F^2.
\end{aligned}
\end{equation}
ADMM then consists of the update steps:
\begin{subequations} \label{RIS:eq21}
\begin{alignat}{1}
& \left(\boldsymbol{\mu}^{i+1}, \mathbf{W}^{i+1}, \mathbf{R}^{i+1}\right)\!=\!\underset{\boldsymbol{\mu}, \mathbf{W}, \mathbf{R}}{\arg \min } \mathcal{L}_\tau\!\!\left(\boldsymbol{\mu}, \mathbf{W}, \mathbf{R}, \mathbf{\Lambda}^i, \mathbf{Z}^i\right),\!\! \label{RIS:eq21a}\\
& \mathbf{Z}^{i+1}=\underset{\mathbf{Z}}{\arg \min } \mathcal{L}_\tau\left(\boldsymbol{\mu}^{i+1}, \mathbf{W}^{i+1}, \mathbf{R}^{i+1}, \mathbf{\Lambda}^i, \mathbf{Z}\right), \label{RIS:eq21b}\\
& \mathbf{\Pi}^{i+1}=\mathbf{\Pi}^i+\tau\left(\mathbf{Z}^{i+1}-\left[\begin{array}{ll}
\mathbf{W}^{i+1} & \left(\mathbf{R}^{i+1}\right)^{\mathrm{H}} \\
\mathbf{R}^{i+1} & \mathbf{T}\left(\boldsymbol{\mu}^{i+1}\right)
\end{array}\right]\right). \label{RIS:eq21c}
\end{alignat}
\end{subequations}
Before updating the steps, we introduced the partitions:
\begin{equation}\label{RIS:eq22}
\mathbf{Z}^i=\left[\begin{array}{cc}
\mathbf{Z}_0^i & \left(\mathbf{Z}_1^i\right)^{\mathrm{H}} \\
\mathbf{Z}_1^i & \mathbf{Z}_2^i
\end{array}\right], \boldsymbol{\Pi}^i=\left[\begin{array}{cc}
\boldsymbol{\Pi}_0^i & \left(\boldsymbol{\Pi}_1^i\right)^{\mathrm{H}} \\
\mathbf{\Pi}_1^i & \boldsymbol{\Pi}_2^i
\end{array}\right],
\end{equation}
and the updates with respect to $\boldsymbol{\mu}$, $\mathbf{W}$, and $\mathbf{R}$ can be computed in closed form:
\begin{equation}\label{RIS:eq23}
\begin{aligned}
& \mathbf{W}^{i+1}= \mathbf{Z}^i_0-\frac{2}{\tau}\left(\mathbf{\Pi}^i_0-\mathbf{I}_N\right), \\
& \mathbf{R}^{i+1}=\frac{1}{\tau+\gamma}{\left(\gamma\widehat{\mathbf{R}}+\mathbf{\Pi}_1^i+\tau\mathbf{Z}_1^i\right)}, \\
& \boldsymbol{\mu}^{i+1}=\boldsymbol{\Lambda}\left\{\mathbf{T}^*\left(\mathbf{Z}_2^i+\mathbf{\Pi}_2^i / \tau\right)-N\mathbf{e}_1/\tau \right\},
\end{aligned}
\end{equation}
where $\mathbf{\Lambda}=\operatorname{diag}\left\{\frac{1}{N}, \frac{1}{2(N-1)}, . ., \frac{1}{2}\right\}$, $\mathbf{e}_1=\left\{1,0,\dots,0\right\}^{\mathrm{T}}$, and $\mathbf{T}^*(\cdot)$ denotes the adjoint operator of $\mathbf{T}(\cdot)$. The $\mathbf{Z}$ update is simply the projection onto the positive definite cone
\begin{equation} \label{RIS:eq24}
\mathbf{Z}^{i+1}=\underset{\mathbf{Z}}{\arg \min }\left\|\mathbf{Z}-\left[\begin{array}{ll}
\mathbf{W}^{i+1} & \left(\mathbf{R}^{i+1}\right)^{\mathrm{H}} \\
\mathbf{R}^{i+1} & \mathbf{T}\left(\boldsymbol{\mu}^{i+1}\right)
\end{array}\right]+\frac{\mathbf{\Pi}^i}{\tau}\right\|_F^2.
\end{equation}
We implement the stopping criteria suggested in \cite{ref13} for our iterative algorithm. Applying MUSIC technique [14] on $\mathbf{T}\left(\boldsymbol{\mu}\right)$, the DOAs
can be finally estimated. The details are shown in Algorithm 2.

\begin{table}[!h]
\normalsize
\centering
\begin{tabular}{p{0.05cm}p{8cm}}
\toprule
   \multicolumn{2}{l}{$\bf{Algorithm~2}$ Gridless DoA Estimation Based on ADMM} \\
\midrule
1. & \emph{Input:} The sample covariance matrix $\widehat{\mathbf{R}}_{\mathbf{Y}}$, the RIS-enabled measurement matrix $\mathbf{B}$, the number of sources $K$, the RIS element location $\mathbf{p}$, the receiving antenna location $\mathbf{q}$, DoD $\alpha$, DoA $\beta$, the maximum number of iterations $I_2$, and estimated noise variance $\hat{\sigma}_0$;\\
2. & \emph{Initialization:} let \emph{i} = 0, and the sample RIS-enabled covariance matrix $\widehat{\mathbf{R}}$ calculated via $(\ref{RIS:eq12})$;\\
3. & $\mathbf{Repeat}$\\
& $\qquad$ $i=i+1$; \\
& $\qquad$ Update the parameter set $\left\{\boldsymbol{\mu}, \mathbf{W}, \mathbf{R}\right\}$ via $(\ref{RIS:eq23})$;\\
& $\qquad$ Update the parameter $\mathbf{Z}$ via $(\ref{RIS:eq24})$;\\
& $\qquad$ Update the parameter $\mathbf{\Pi}$ via $(\ref{RIS:eq21c})$;\\
& $\mathbf{Until}~i>I_2$ or halting criteria suggested in \cite{ref13};\\
4. & Apply MUSIC method on $\mathbf{T}\left(\boldsymbol{\mu}\right)$ to get the DoAs;\\
5. & \emph{Output:} estimated DoAs $\widehat{\theta}_k$.\\
\bottomrule
\end{tabular}
\end{table}

\section{Simulation}

In this section, we illustrate the performance of our proposed approach by simulations and compare it with the ANM-based algorithm in the time domain [11]. All simulations are collected through 100 Monte-Carlo trails. To demonstrate the effectiveness of the proposed method in the NLoS situation, we choose the root mean square error (RMSE) and CPU time as metrics.

Neither the proposed method nor the baseline involves any optimization of the phased matrix of RIS, hence the phase entry of RIS is only randomly assigned the two values of 0$^{\circ}$ or 180$^{\circ}$. Not otherwise specified, assume three sources impinge onto RIS from distinct DoAs of $\left[5.345^\circ,25.789^\circ,45.456^\circ\right]$ and fix $M=4$ isotropic antennas of BS for receiving data.

\begin{figure}[!htb]
	\centering
	\includegraphics[width=0.71\linewidth]{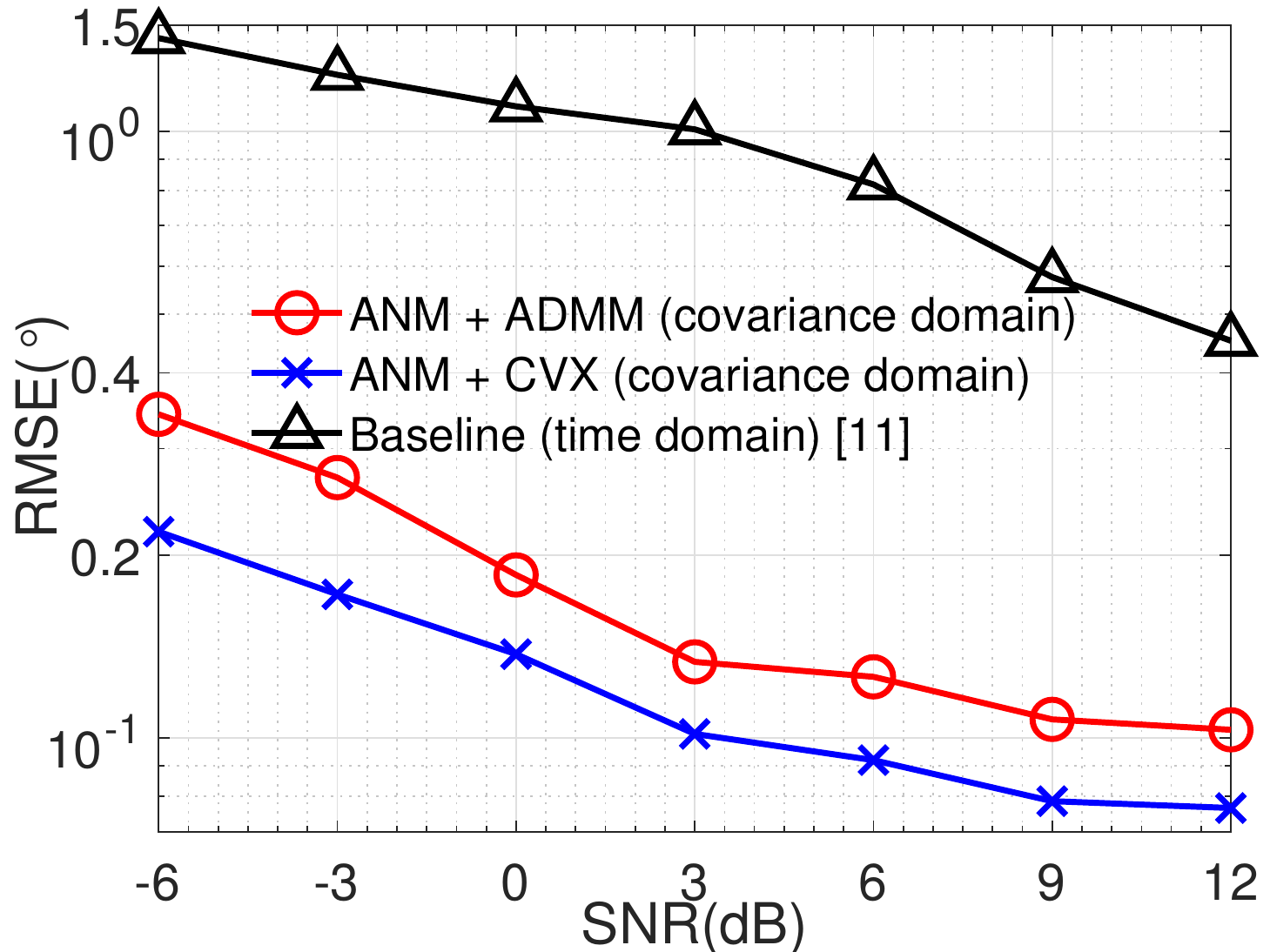}
	\caption{RMSE curve versus SNR with $L=32$ and $N=16$\protect\footnotemark.}
	\label{fig:snr}
\end{figure}
\footnotetext{Since the baseline will produce misjudgments under non-ideal conditions such as low SNR, limited RIS elements, and limited number of measurements, we set the estimation error to 4 degrees when the true estimation error based on the baseline is greater than 4 degrees for better illustration.}

\begin{figure}[!htb]
	\centering
	\includegraphics[width=0.71\linewidth]{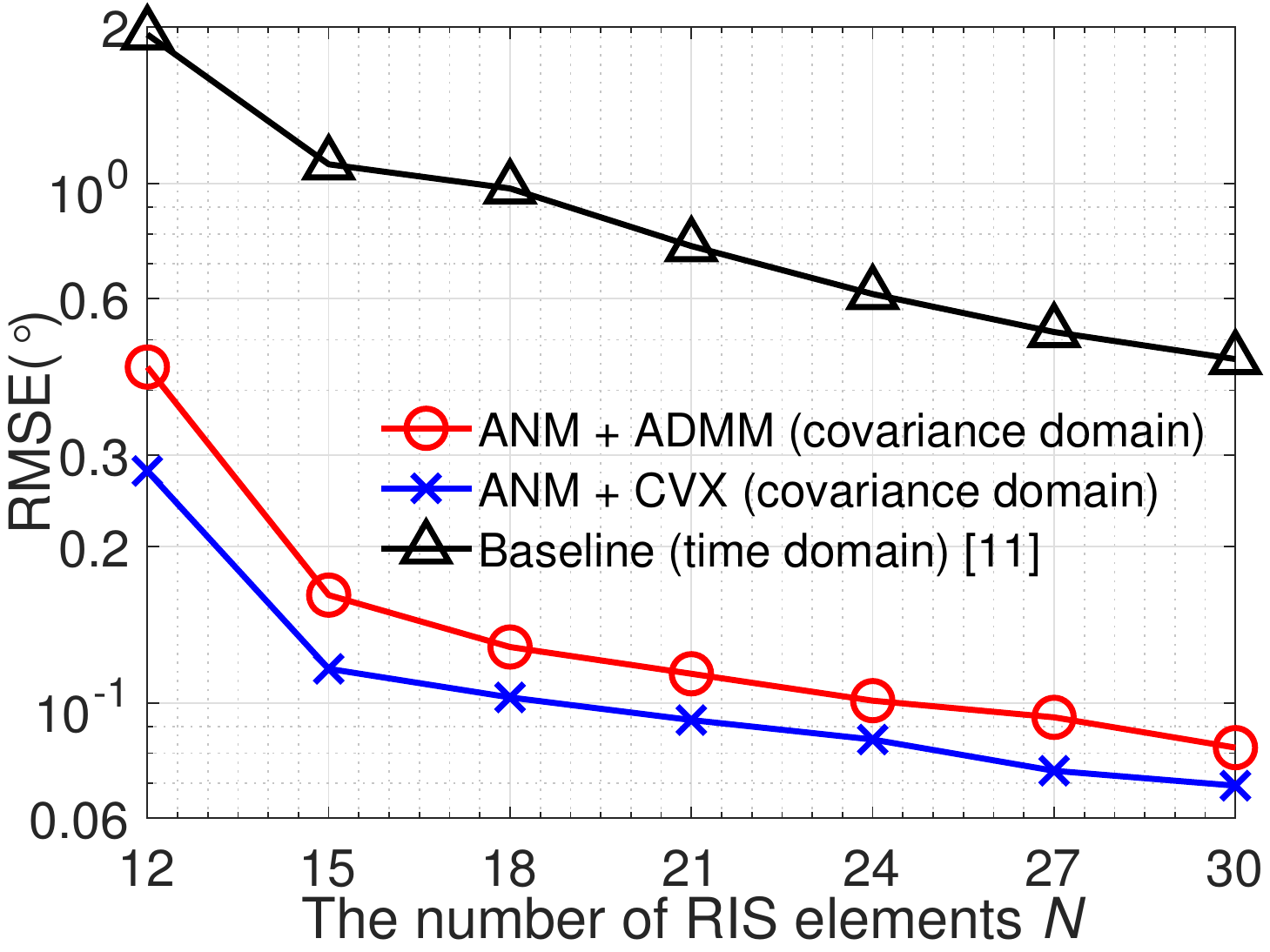}
	\caption{RMSE curve versus RIS elements with SNR $=3$ dB and $L=32$.}
	\label{fig:ris}
\end{figure}
We compare the accuracy of multi-source DoA estimation under various algorithms in terms of the signal-to-ratio (SNR) conditions, the number of RIS elements, and the number of measurements. In the first experiment, Fig.2 depicts the RMSE curve as SNR varies from -6 dB to 12 dB with $L=32$ and $N=16$. It is observed that the proposed method is robust in low SNR conditions compared to the benchmark. This is attributed both to the excellent denoising operation that exploits the noise statistics and to the salient recovery performance of the gridless CS framework. It is also clear that the interior point method embedded in the CVX toolbox \cite{ref12} boils down to the CVX toolbox typically outperforming the ADMM algorithm.

We fix the parameters such as SNR$ =3$ dB and $L=32$, and vary the number of RIS elements from 12 to 30 in the second example. It is seen that the time domain approach degrades dramatically due to the limited RIS elements, and even fails in the DoA estimation in Fig.3. In particular, the resulting RMSE in baseline exceeds two degrees when $N=12$ . However, the proposed algorithm performs well throughout the whole element area.

\begin{figure}[!htb]
	\centering
	\includegraphics[width=0.71\linewidth]{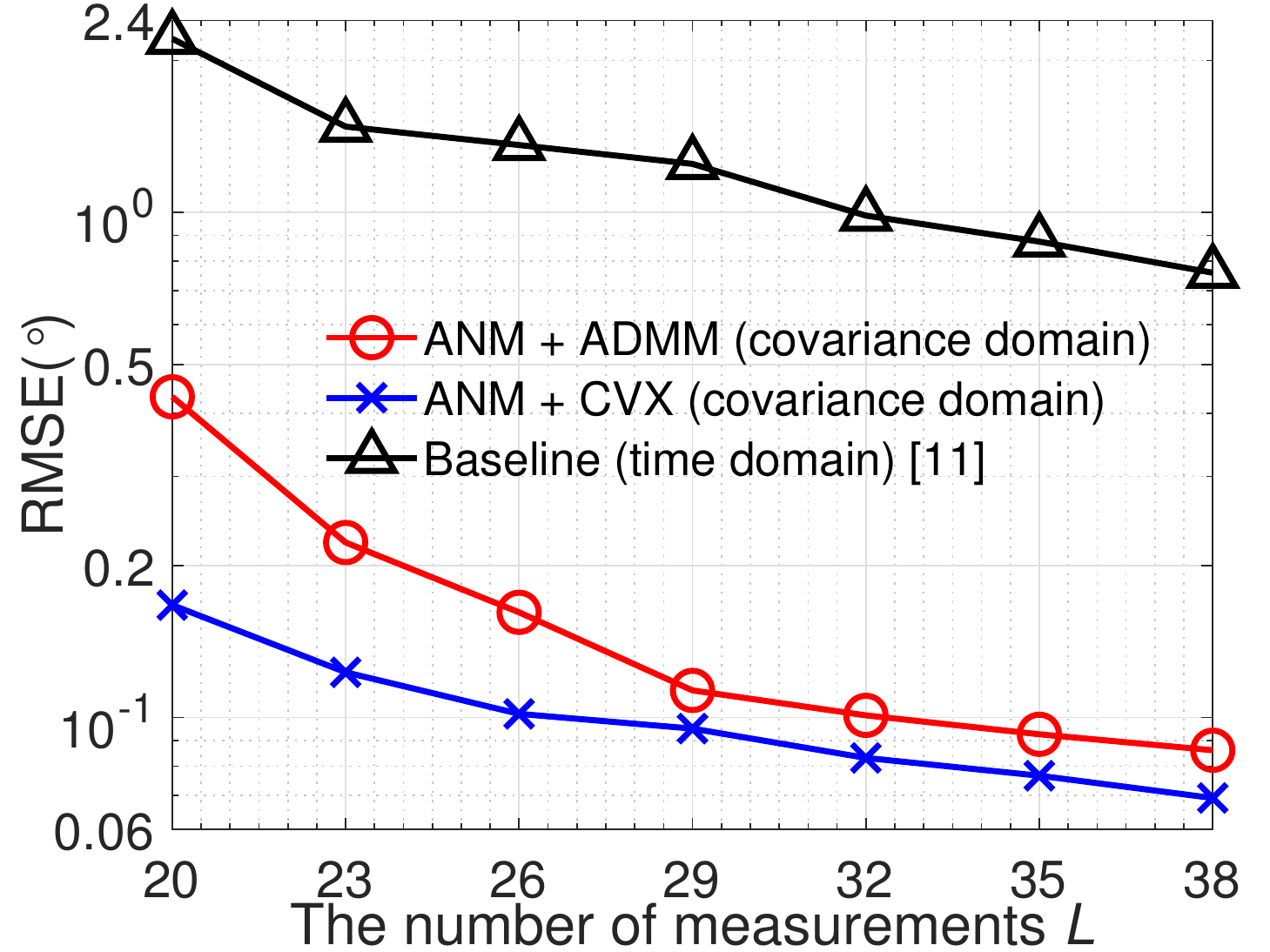}
	\caption{RMSE curve versus measurements with SNR $=3$ dB and $N=16$.}
	\label{fig:snap}
\end{figure}
Now, we vary the number of measurements from 20 to 38 while fixing SNR = 3 dB and $N=16$. Fig.4 plots the RMSE curve to demonstrate the estimation accuracy with different measurements. Similar to the previous simulation, the resulting RMSEs reflect the proposed method substantially outperforms the benchmark. Also, based on our observations in the above three examples, the performance of resulting RMSEs disparity widens as these argument parameters are reduced. This fully demonstrates the superiority of the proposed algorithm under non-ideal conditions such as limited RIS elements, limited number of measurements, and low SNR cases.

\begin{figure}[!htb]
	\centering
	\includegraphics[width=0.71\linewidth]{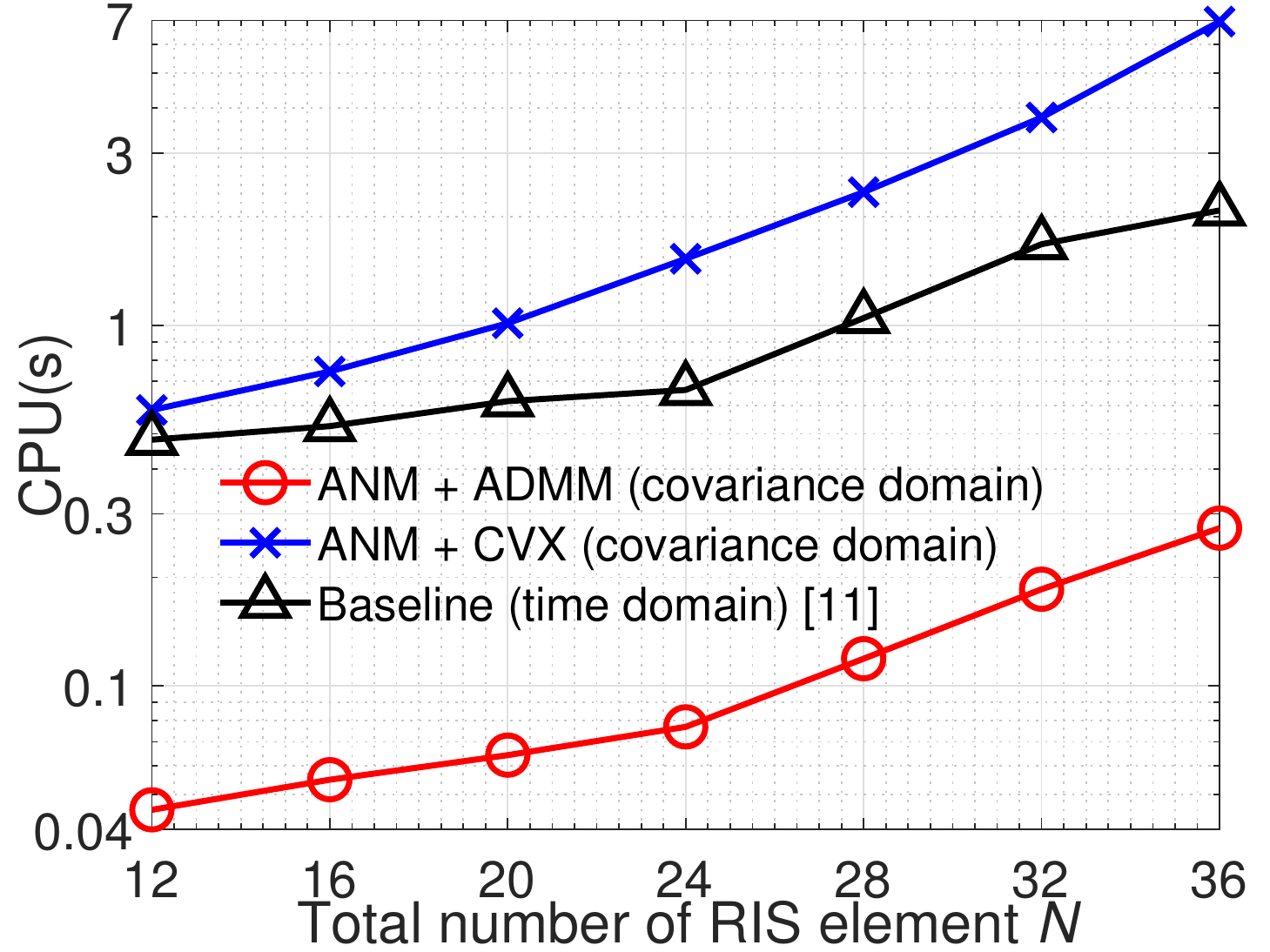}
	\caption{CPU time versus RIS elements with SNR $=3$ dB and $L=36$.}
	\label{fig:cpu}
\end{figure}
To evaluate the computation complexity of different algorithms, we consider the CPU time in various RIS elements with SNR $=$ 3 dB and $L=36$ and plot the results in Fig.5. Specifically, it is supposed that the ANM-based algorithm processes each antenna in parallel in the time domain, which accounts for its operating speed advantage over the proposed algorithm in the covariance domain via the CVX toolbox.

\section{Conclusion}
To perform DoA estimation in NLoS scenarios, this paper has presented a gridless DoA estimation algorithm based on the RIS to generate high-precision estimators of multi-source DoAs. The noise variance is determined by minimizing the Frobenius norm of the measurement error matrix in the covariance domain, and the denoising operation is carried out on the RIS-enabled covariance matrix. CVX toolkit solves the DoA estimation issue by converting it into the ANM problem in a RIS-enabled covariance matrix. Due to the sluggish convergence rate of the CVX toolkit, an ADMM-based efficient iterative algorithm is developed. Simulation results show that the proposed approach provides an improved DOA estimation with lower computational complexity.

\end{document}